\documentclass[english,ngerman]{bvm}

\bvmdef\articlenumber{0000}

\bvmdef\type{V}

\date{}

\usepackage{fancyhdr}
\fancypagestyle{specialfooter}{%
  \fancyhf{}
  
  \fancyfoot[L]{\scriptsize{$\copyright$ Bildverarbeitung f\"ur die Medizin Workshop 2023 (BVM 2023), scheduled for 2-4 July 2023 in Braunschweig, Germany.}}
}														 					  

\title{A Vessel-Segmentation-Based CycleGAN for Unpaired Multi-modal Retinal Image Synthesis}

\titlerunning{Retinal Image Synthesis}

\author{Aline Sindel, Andreas Maier, Vincent Christlein}

\institute{Pattern Recognition Lab, FAU Erlangen-N\"urnberg}

\email{aline.sindel@fau.de}

\begin{document}
						   
\pagestyle{myheadings}

%
\selectlanguage{english}

\maketitle
\thispagestyle{specialfooter}

\begin{abstract}
Unpaired image-to-image translation of retinal images can efficiently increase the training dataset for deep-learning-based multi-modal retinal registration methods. Our method integrates a vessel segmentation network into the image-to-image translation task by extending the CycleGAN framework. The segmentation network is inserted prior to a UNet vision transformer generator network and serves as a shared representation between both domains. We reformulate the original identity loss to learn the direct mapping between the vessel segmentation and the real image. Additionally, we add a segmentation loss term to ensure shared vessel locations between fake and real images. In the experiments, our method shows a visually realistic look and preserves the vessel structures, which is a prerequisite for generating multi-modal training data for image registration.
\end{abstract}

\section{Introduction}
Recent deep learning methods for multi-modal medical image registration require a large amount of training data. Since it is difficult and tedious to obtain precise ground truth from real data, image-to-image translation methods are effective means to synthetically augment multi-modal datasets. In ophthalmology, different imaging systems, such as color fundus (CF), fluorescein angiography (FA), and optical coherence tomography angiography (OCTA) are used for the diagnosis of retinal diseases. Our aim is to generate synthetic multi-modal pairs that can be used to train registration methods in a self-supervised manner. For that the position and shape of the vessels should be preserved by the image-to-image translation method, but the texture and style should be transferred to the other modality. Here, we concentrate on the image-to-image translation between CF and FA images. In CF the vessels are depicted in dark and in FA in light, but by both modalities the fovea is depicted in dark and the optic cup and disc are depicted in light, which needs to be considered by the translation methods.

Conditional generative adversarial networks (cGANs) were explored in literature for the image-to-image translation of CF and FA images. In case of aligned multi-modal images, Pix2Pix~\cite{IsolaP2017} based approaches can be used to learn a direct 1-to-1 mapping between both modalities. In this regard, VTGAN~\cite{KamranSA2021} is introduced for closely but not perfectly aligned CF-FA image pairs, which uses a coarse and a fine generator with attention blocks and a vision transformer as discriminator network. 
CycleGAN~\cite{ZhuJY2017} based approaches learn a direct image-to-image translation for unpaired images. 
For the CF-FA translation task, Li et al.~\cite{LiK2019} enriches the CycleGAN with structure and appearance encoder networks which are inserted prior to the generator networks and Cai et al.~\cite{CaiZ2020} extends the CycleGAN with multi-scale generator and discriminator networks and a quality-aware loss at feature level.
In contrast, we extend the CycleGAN by including the vessel segmentation as a shared representation between both domains.
There exist a bunch of approaches that generate CF images from extracted vessel segmentations. 
For instance, the cGAN by Liang et al.~\cite{LiangN2022} adds a class feature loss for diabetic retinopathy grading and a retinal detail loss which is a combination of the reconstruction loss between real and fake image and perceptual loss using a specific layer from the pretrained VGG-19 network. Niu et al.~\cite{NiuY2022} includes pathology specific descriptors into the cGAN to generate CF images with specific pathological features. The real and synthetic images are compared using perceptual and severity losses. 
By integrating the vessel segmentation into the CycleGAN, we tackle to reduce the domain gap of the vessels between both modalities.

In this paper, we propose VesselCycleGAN, a cGAN based approach for unpaired retinal image-to-image translation based on the vessel segmentation of CF and FA images using cycle consistency. We extend the CycleGAN pipeline by inserting a vessel segmentation UNet before the generator network, which we equip with a UNet vision transformer~\cite{TorbunovD2022}. With the vessel segmentation network, we modify the identity loss to learn the translation from the vessel segmentation to the real image and we add a segmentation loss to preserve the same vessel structures in the real and generated images and apply our method to two datasets.

\begin{figure}[t]
	\setlength{\figbreite}{0.8\textwidth}
	\centering
	\caption{Our unpaired image-to-image translation method based on the retinal vessel segmentation in multi-modal fundus images.}
	\includegraphics[width=\figbreite]{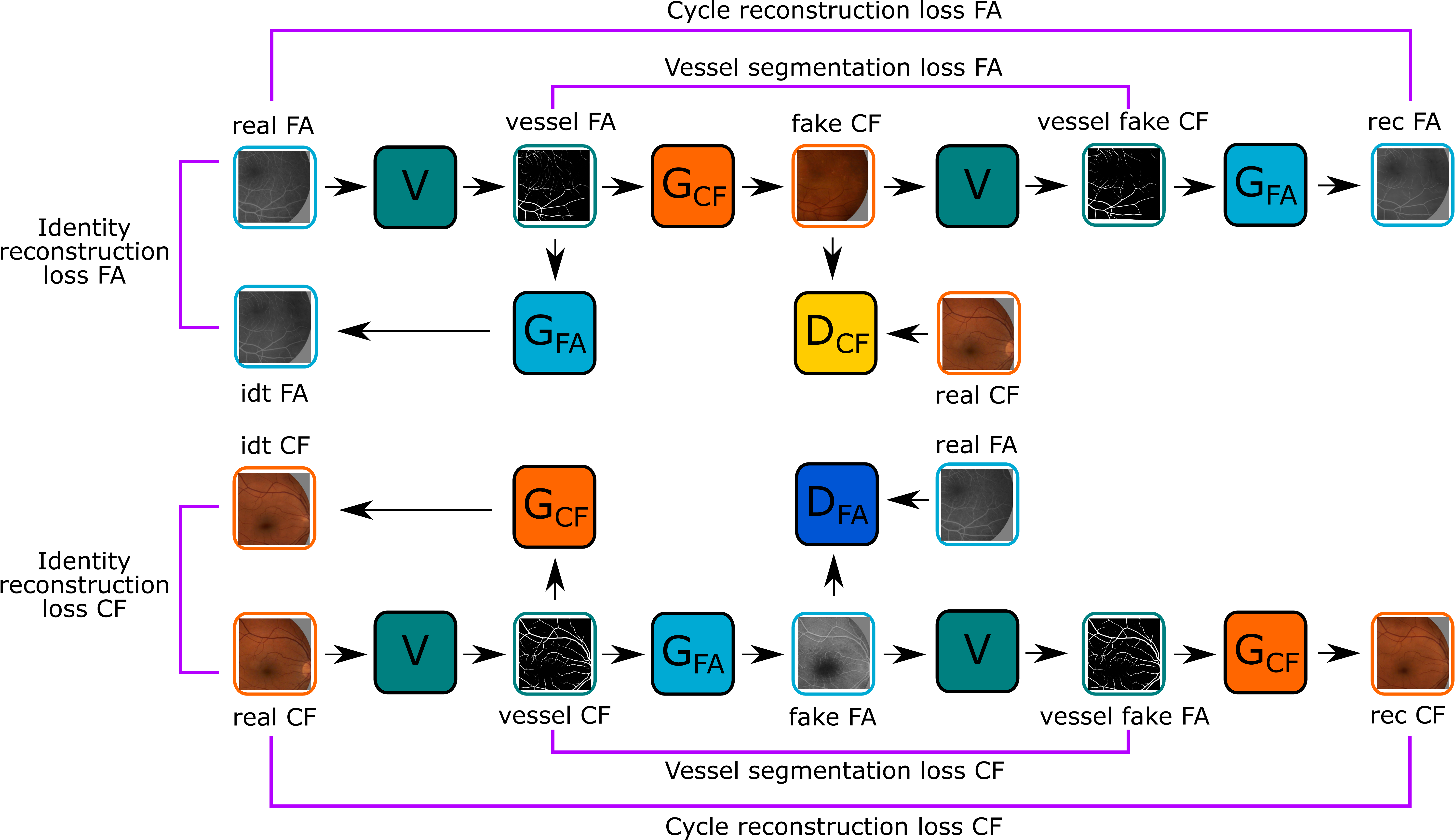}
	\label{fig-01}
\end{figure} 

\section{Materials and methods}
\subsection{VesselCycleGAN for unpaired retinal image-to-image translation}
We incorporate a vessel segmentation network (V) into the CycleGAN~\cite{ZhuJY2017} framework, as shown in Figure~\ref{fig-01}, which we place in front of the generator networks, such that those do not learn the direct mapping between the two domains, but the mapping from the vessel segmentation to the particular domain. 
CycleGANs are conditional generative adversarial networks, consisting of two generator $G_{A/B}$ and two discriminator networks $D_{A/B}$, that learn the image-to-image translation between unpaired images from two domains $A/B$ by using adversarial loss, cycle-consistency and identity-consistency losses~\cite{ZhuJY2017}. The cycle-consistency loss $\mathcal{L}_{\text{cycA}}$ minimizes the difference between the real image $A$ and its reconstruction $\hat{A}$ after passing through the cycle of applying both $G_B$ and $G_A$, and here in our case $V$, $G_B$, $V$, and $G_A$:
\begin{equation}
\mathcal{L}_{\text{cycA}}(A) = \lambda_A ||A - G_A(V(G_B(V(A))))||_1. 
\end{equation}
Using the identity-consistency loss $\mathcal{L}_{\text{idtA}}$, $G_A$ originally learns the identity mapping of $G_A(A) \overset ! = A$, however, in our modified CycleGAN this becomes:
\begin{equation}
\mathcal{L}_{\text{idtA}}(A) = \lambda_A \lambda_{idt} ||A - G_A(V(A))||_1, 
\end{equation}
where $G_A$ learns the aligned one-way image translation task of vessel segmentation $A$ to real $A$. Additionally, we compute the Dice loss between the segmentation of the real $A$ and fake $B$. The role of $D_{A/B}$ is as in the normal CycleGAN to distinguish unpaired real images $A/B$ from fake images $\tilde{A}/\tilde{B}$ generated using $G_A(V(B))$ or $G_B(V(A))$.
As generator network we employ the UNet-ViT~\cite{TorbunovD2022} which is a four layer UNet ($D=48$) with a pixel-wise vision transformer bottleneck (with $12$ transform encoder blocks). As discriminator network we use the PatchGAN~\cite{IsolaP2017} and for the vessel segmentation network a four layer UNet ($D=16$), which we pretrained for the vessel segmentation task.

\subsection{Retinal datasets}
We train and test our synthesis method using the CF-FA dataset~\cite{ShirinH2021} which consists of $59$ pairs of color fundus (CF, $576 \times 720$) and fluorescein angiography (FA, $576 \times 720$) images from controls ($29$ image pairs) and from patients with diabetic retinopathy ($30$ pairs).
We split the image pairs into train: $35$, val: $10$, and test: $14$, with equally distributed healthy and non-healthy eyes.
For each training image, we extract nine $512 \times 512$ patches, which are randomly cropped to $448 \times 448$. For the validation and test set, we directly extract up to nine $448 \times 448$ patches for each image. This results per modality into train: $315$, val: $90$ and test: $120$ image patches. Since the original image pairs are not aligned, we register the images of the test set (prior to patch extraction) using  KPVSA-Net~\cite{SindelA2022}.

Secondly, we use the HR fundus dataset~\cite{BudaiA2013} which consists of CF images and manual vessel segmentations. 
To train the vessel segmentation UNet, we use the color and green channel of the fundus images in two resolutions: the center $768 \times 768$ region and a downsized version by a factor of $4$ to have a similar image size as the CF-FA dataset. During training, we randomly extract $512 \times 512$ patches on the fly from the $108$ training and $36$ validation images.
For the image translation task, we use $81$ $484 \times 484$ patches from the CF images of the test set without the ground truth vessel segmentations.

\begin{figure}[b]
	\setlength{\figbreite}{0.81\textwidth} 
	\centering	
	\includegraphics[width=\figbreite]{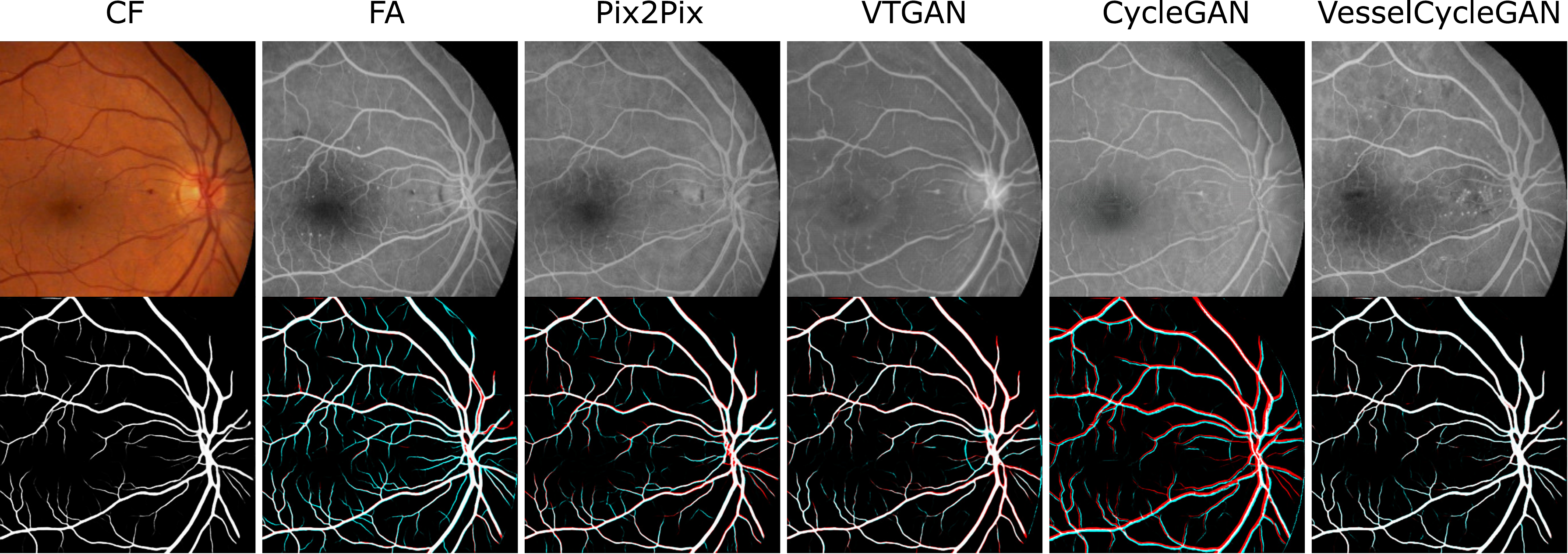}
	\caption{FA synthesis results and vessel segmentation overlays (vessels segmented by our UNet).}		
	\label{fig-02}
\end{figure}

\begin{figure}[b]
	\setlength{\figbreite}{0.68\textwidth} 
	\centering
	\includegraphics[width=\figbreite]{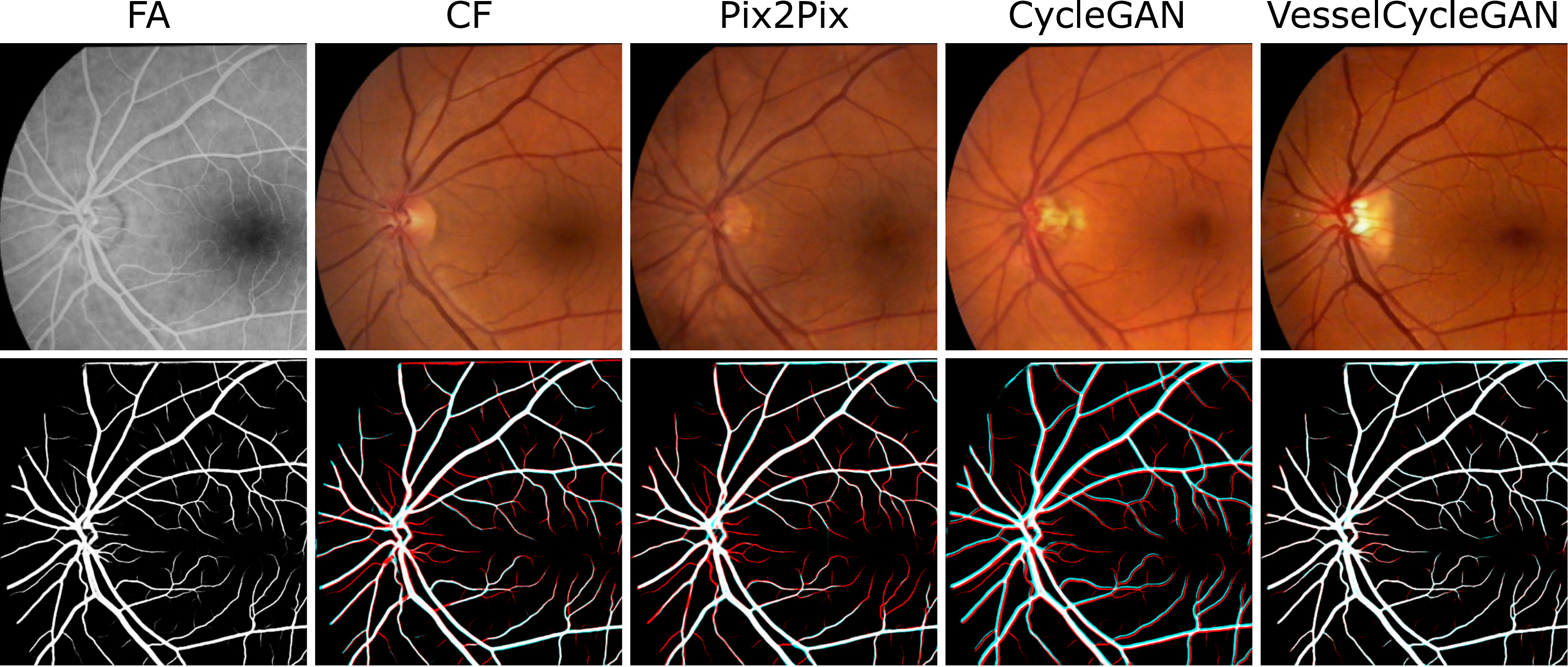}	
	\caption{CF synthesis results and vessel segmentation overlays (vessels segmented by our UNet).}
	\label{fig-03}
\end{figure}

\begin{figure}[b]
	\setlength{\figbreite}{0.68\textwidth} 
	\centering	
	\includegraphics[width=\figbreite]{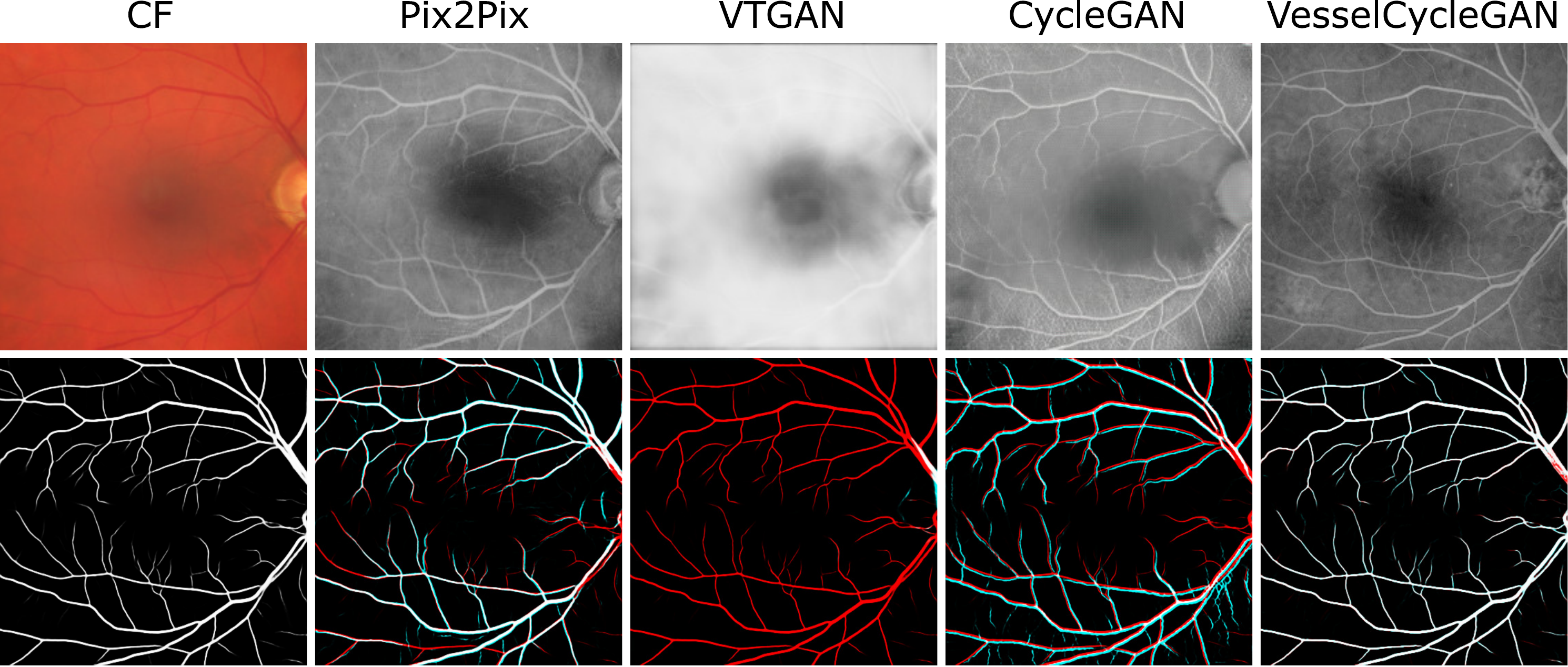}	
	\caption{FA synthesis results of the HRF dataset (vessels segmented by our UNet).}
	\label{fig-04}
\end{figure}

\subsection{Experimental details}
Prior to the training of our retinal synthesis network, we train the vessel segmentation UNet on the augmented HR fundus dataset by using equally weighted binary cross-entropy and Dice loss for $800$ epochs with early stopping, Adam optimizer, a learning rate $\eta=2\cdot10^{-4}$, linear decay of $\eta$ after $50$ epochs, and batch size of $2$.
Then, we train the generator and discriminator networks of our retina synthesis GAN using Adam solver with a learning rate of $\eta=2\cdot10^{-4}$ for $600$ epochs with early stopping, a batch size of $1$, $\lambda_\text{A/B}=100$, $\lambda_\text{idt}=1$,  $\lambda_\text{seg}=1$.
For both tasks, we use online data augmentation (color jittering, horizontal flipping, rotation, and cropping). 
We compare our method with CycleGAN~\cite{ZhuJY2017} and Pix2Pix~\cite{IsolaP2017} (both: $G$ using ResNet encoder with $9$ blocks with instance normalization), and with the VTGAN~\cite{KamranSA2021}.
Pix2Pix and VTGAN require aligned training data, hence we registered our training and validation image pairs using KPVSA-Net~\cite{SindelA2022}.
For our method and CycleGAN, we use unaligned data with randomly sampled patches within the same class (healthy/unhealthy).

\begin{table}[t]
\caption{Quantitative evaluation for FA and CF synthesis using the CF-FA test images. LPIPS and KID are computed between fake B and registered real B; Dice between the vessels of fake B and real A. Methods with * were trained using registered images.} 
\label{tab-01}
\begin{tabular*}{\textwidth}{l@{\extracolsep\fill}cccccc}
\hline
A-to-B & \multicolumn{3}{c}{CF-to-FA} & \multicolumn{3}{c}{FA-to-CF} \\
\cline{2-4}
\cline{5-7}
Metrics & LPIPS $\downarrow$ & KID $\downarrow$ &  Dice $\uparrow$  & LPIPS $\downarrow$ & KID $\downarrow$ & Dice $\uparrow$ \\
\hline
Pix2Pix* (ResNet9) & \emph{0.3478} & \emph{0.0104} & 0.7615 & 0.3371 & 0.0227 & 0.7264 \\ 
VTGAN* & 0.3731 & 0.0383 & 0.8035 & - & - & - \\ 
CycleGAN (ResNet9) & 0.4296 & 0.0388 & 0.2471 & 0.3511 & 0.0182 & 0.3550 \\ 
VesselCycleGAN (ResNet9 w/o seg) & 0.3893 & 0.0322 & 0.8798 & 0.3507 & 0.0211 & 0.8458 \\ 
VesselCycleGAN (w/o seg) & 0.3652 & 0.0158 & 0.8471 & 0.3503 & 0.0227 & 0.8567 \\ 
VesselCycleGAN & 0.3685 & 0.0163 & \emph{0.9534} & \emph{0.3329} & \emph{0.0148} & \emph{0.9419} \\
\hline
\end{tabular*}
\end{table}

\begin{table}[t]
\caption{Quantitative evaluation for FA synthesis for the HRF test images using the models trained on the CF-FA dataset. For KID, real FAs are from the CF-FA dataset, since there are none in the HRF dataset. Methods with * are trained using registered CF-FA images.} 
\label{tab-02}
\begin{tabular*}{\textwidth}{l@{\extracolsep\fill}llll}
\hline
Metrics & Pix2Pix* & VTGAN* & CycleGAN & VesselCycleGAN \\ 
\hline
KID $\downarrow$ (unaligned fake - real FA) & \emph{0.0295} & 0.2010 & 0.0580 & 0.0515 \\ 
Dice  $\uparrow$ (fake FA - real CF)& 0.6943 & 0.3211 & 0.2502 & \emph{0.9439} \\ 
\hline
\end{tabular*}
\end{table}

\section{Results}
Table~\ref{tab-01} summarizes the quantitative results for the CF-to-FA and FA-to-CF task using similarity (LPIPS, KID) and Dice metrics. For the FA synthesis, Pix2Pix obtained the best LPIPS and KID scores, but has only a relatively low Dice score of $0.76$. Our VesselCycleGAN achieves a bit lower similarity scores, but the highest Dice score of $0.95$ and is superior to VTGAN and the default CycleGAN.
For the CF synthesis task, our VessselCycleGAN achieves the best LPIPS, KID, and Dice scores. Pix2Pix here only achieves the second best LPIPS score. 
For both synthesis tasks, the Dice scores of CycleGAN are very low, indicating that the vessel positions have not been preserved in the synthetic image.
Moreover, we tested different settings for our VesselCycleGAN to show the advantage of adding the segmentation loss (gain of up to $10$ \% Dice score) and by using the UNet-ViT instead of the ResNet9 generator network.
The qualitative results in Figure~\ref{fig-02} and ~\ref{fig-03} reflect the findings. The structures in the synthetic images generated by VesselCycleGAN and Pix2Pix demonstrate visual similar structures to the real images. In the vessel segmentation overlays, deviating vessels between the content and generated image are marked in red (missed vessels) and cyan (added vessels). Our VesselCycleGAN has the highest overlap in the vessel structure with the content image. Pix2Pix and VTGAN show some small misalignment in the vessel details, as they learn a direct mapping between the domains from the registered data, which can show some small deformations in the vessel structure.
Further, we tested the transferability of the trained models for CF images of the HRF dataset, where no ground truth FA images exist. In Figure~\ref{fig-04}, the synthetic image of VesselCycleGAN depicts the fine structures, the result of Pix2Pix is a bit blurry, very blurry for CycleGAN while VTGAN was not able to obtain a realistic result. Numerically, the KID between the FA images from the CF-FA dataset and the generated FA images of the CF HRF images in Table~\ref{tab-02}, was best for Pix2Pix and second best for VesselCycleGAN. The Dice score for VesselCycleGAN is close the CF-FA dataset, while the competing methods have lower or very low results.

\section{Discussion}
We relax the unpaired image-to-image translation of multi-modal fundus images to direct mappings from the vessel segmentation to the other modality within the CycleGAN pipeline. Our method, which is trained with unpaired images, learns the modality specific visual patterns and preserves the vessel locations, and thus can be used to augment training data for multi-modal retinal registration methods. 
As future work, our vessel-based approach could be extended by also including optical disc segmentations and further methods to control the synthesis of pathological structures could be investigated.

\bibliographystyle{bvm}

\bibliography{ref}

\begin{thebibliography}{10}

\bibitem{IsolaP2017}
Isola P, Zhu JY, Zhou T, et~al.
\newblock {Image-To-Image Translation With Conditional Adversarial Networks}.
\newblock Proc IEEE CVPR 2017. 2017;.

\bibitem{KamranSA2021}
Kamran SA, Hossain KF, Tavakkoli A, et~al.
\newblock {VTGAN: Semi-supervised Retinal Image Synthesis and Disease
  Prediction using Vision Transformers}.
\newblock 2021 IEEE/CVF ICCVW. 2021; p. 3228--3238.

\bibitem{ZhuJY2017}
Zhu JY, Park T, Isola P, et~al.
\newblock {Unpaired Image-to-Image Translation using Cycle-Consistent
  Adversarial Networks}.
\newblock Proc IEEE ICCV 2017. 2017; p. 2242--2251.

\bibitem{LiK2019}
Li K, Yu L, Wang S, et~al.
\newblock {Unsupervised Retina Image Synthesis via Disentangled Representation
  Learning}.
\newblock SASHIMI 2019. 2019; p. 32--41.

\bibitem{CaiZ2020}
Cai Z, Xin J, Wu J, et~al.
\newblock {Triple Multi-scale Adversarial Learning with Self-attention and
  Quality Loss for Unpaired Fundus Fluorescein Angiography Synthesis}.
\newblock IEEE EMBC 2020. 2020; p. 1592--1595.

\bibitem{LiangN2022}
Liang N, Yuan L, Wen X, et~al.
\newblock {End-To-End Retina Image Synthesis Based on CGAN Using Class Feature
  Loss and Improved Retinal Detail Loss}.
\newblock IEEE Access. 2022;10:83125--83137.

\bibitem{NiuY2022}
Niu Y, Gu L, Zhao Y, et~al.
\newblock {Explainable Diabetic Retinopathy Detection and Retinal Image
  Generation}.
\newblock IEEE J Biomed Health Inform. 2022;26(1):44--55.

\bibitem{TorbunovD2022}
Torbunov D, Huang Y, Yu H, et~al.
\newblock {UVCGAN: UNet Vision Transformer Cycle-Consistent GAN for Unpaired
  Image-to-Image Translation}.
\newblock Proc IEEE/CVF WACV 2023. 2023; p. 702--712.

\bibitem{ShirinH2021}
Hajeb Mohammad~Alipour S, Rabbani H, Akhlaghi MR.
\newblock {Diabetic Retinopathy Grading by Digital Curvelet Transform}.
\newblock Comput Math Methods Med. 2012;2021:761901.

\bibitem{SindelA2022}
Sindel A, Hohberger B, Maier A, et~al.
\newblock {Multi-modal Retinal Image Registration Using a Keypoint-Based Vessel
  Structure Aligning Network}.
\newblock MICCAI 2022. 2022; p. 108--118.

\bibitem{BudaiA2013}
Budai A, Bock R, Maier A, et~al.
\newblock {Robust Vessel Segmentation in Fundus Images}.
\newblock Int J Biomed Imaging. 2013;.

\end{thebibliography}

\marginpar{\color{white}E\articlenumber}			  

\end{document}